\documentclass[12pt,preprint]{aastex}

\newcommand{\bllac}{{BL\,Lac}}
\newcommand{\pcal}{J2153+4322}

\def\arcdeg{$^\circ$}

\shortauthors{Dodson, et al.}
\shorttitle{High-precision Astrometric mm-VLBI Using a New Method for Multi-Frequency Calibration}

\title{High-precision Astrometric Millimeter Very Long Baseline Interferometry Using a New Method for Multi-Frequency Calibration}

 \author{
   Richard \textsc{Dodson}\altaffilmark{1},
   Mar\'{\i}a J. \textsc{Rioja}\altaffilmark{1,2,3},
   Sol N. \textsc{Molina}\altaffilmark{4},
   Jos\'{e} L. \textsc{G\'{o}mez}\altaffilmark{4}
}
 \affil{$^1$ International Centre for Radio Astronomy Research, M468, The University of Western Australia, 
             35 Stirling Hwy, Crawley, Western Australia, 6009}
 \affil{$^2$ CSIRO Astronomy and Space Science, 26 Dick Perry Avenue, Kensington WA 6151, Australia}
 \affil{$^3$ Observatorio Astron\'omico Nacional (IGN), Alfonso XII, 3 y 5, 28014 Madrid, Spain} 
 \affil{$^4$ Instituto de Astrof\'{\i}sica de Andaluc\'{\i}a-CSIC, Glorieta de la Astronom\'{\i}a s/n, E-18008 Granada, Spain}
 \email{richard.dodson@icrar.org}


\begin{document}

\begin{abstract} 
In this paper we describe a new approach for mm-VLBI calibration that provides bona-fide astrometric alignment of the mm-wavelength images from a single source, for the measurement of frequency dependent effects, such as `core-shifts' near the black hole of AGN jets. We achieve our astrometric alignment by solving firstly for the ionospheric (dispersive) contributions using wide-band cm-wavelength observations. { Secondly we solve for the tropospheric (non-dispersive) contributions by using fast frequency-switching at the target mm-wavelengths. These solutions can be scaled and transferred} from the low frequency to the high frequency. To complete the calibration chain one additional step was required to remove a residual constant phase offset on each antenna. The result is an astrometric calibration and the measurement of the core-shift between 22 and 43 GHz for the jet in BL Lacertae to be -8$\pm$5, 20$\pm$6 $\mu$as, in RA and Declination, respectively. By comparison to conventional phase referencing at cm-wavelengths we are able to show that this core shift at mm-wavelengths is significantly less than what would be predicted by extrapolating the low frequency result, which closely followed the predictions of the Blandford \& K\"onigl conical jet model.
As such it would be the first demonstration for the association of the VLBI core with a recollimation shock, normally hidden at low frequencies due to the optical depth, which could be responsible for the $\gamma$-ray production in blazar jets.
\end{abstract}
\maketitle
\section{Introduction}\label{sec:intr}

{ Results of over eight years of monthly monitoring of a sample of blazars (the most luminous and variable BL~Lac objects and flat-spectrum radio quasars) with the VLBA at 7~mm by the Boston University blazar group \citep{JorstadMarscher2016}\footnote{The VLBA-BU-BLAZAR monitoring program; see http://www.bu.edu/blazars/VLBAproject.html} show that most $\gamma$-ray flares are simultaneous (within errors) with the appearance of a new superluminal component or a major outburst in the VLBI core of the jet, defined as the bright, compact feature at the upstream end of the jet \citep[see][]{marscher_08,marscher_10,2013ApJ...773..147J,2015ApJ...808..162C,2015ApJ...813...51C}. A burst in particle and magnetic energy density is therefore required when jet disturbances cross the radio core in order to produce $\gamma$-ray flares, which can naturally be explained by identifying the radio core with a recollimation shock \citep[e.g.,][]{daly_88,gomez_95,gomez_97,Marscher:2009vs,Marscher:2012gh,2015ApJ...809...38M,Marti2016}.}

On the other hand, the standard Blandford \& K\"onigl conical jet model hypothesizes that the core is not a physical feature in the jet, but corresponds to the location at which the jet becomes optically thin, and therefore its position shifts with observing frequency \citep{bk_79,konigl_81,lob_98}. This is conventionally referred to as the `core-shift'. In this case the separation from the black hole is $r$=$r_0 \nu^\kappa$, where $\nu$ is the frequency and $\kappa$ is a value close to -1 \citep[][additionally there can be an offset from the nominal reference position]{lob_98}. Multi-frequency VLBI observations at centimeter wavelengths have measured this core frequency shift in multiple sources, albeit without phase-referencing \citep[e.g.,][]{kovalev_08,osullivan_09,sokolovsky_11,fromm_15}. {\bf Nevertheless phase-referenced VLBI observations have confirmed that the cm-wavelength radio core indeed is consistent with the optically thick-thin transition, in a smaller number of targets, such as 3C\,395, 4C\,39.25, 1038+528, 3C\,390.1, M\,81, M\,87 and 3C454.3 \citep[respectively]{lara_94,guirado_95,rioja_98,ros_01,martividal_11,hada_11,kutkin_14}.}

We have therefore two sets of results, one suggesting that the radio core corresponds to a recollimation shock while the other implies that it marks the transition between the optically thick-thin jet regimes. A possible solution to reconcile these apparently contradicting observational results is to consider that the core is located parsecs away from the central black hole \citep[e.g.,][]{marscher_02,chatterjee_11,fromm_15} and consists of a recollimation shock that leads to $\gamma$-ray flares as new perturbations in the jet flow cross its position \citep[e.g.,][]{JorstadMarscher2016}. At this distance from the black hole the core is optically thin at mm-wavelengths, while at longer wavelengths the core becomes optically thick, leading to the observed Blandford \& K\"onigl core frequency shift.

\begin{figure}[t]
\begin{center}
\includegraphics[width=0.32\textwidth]{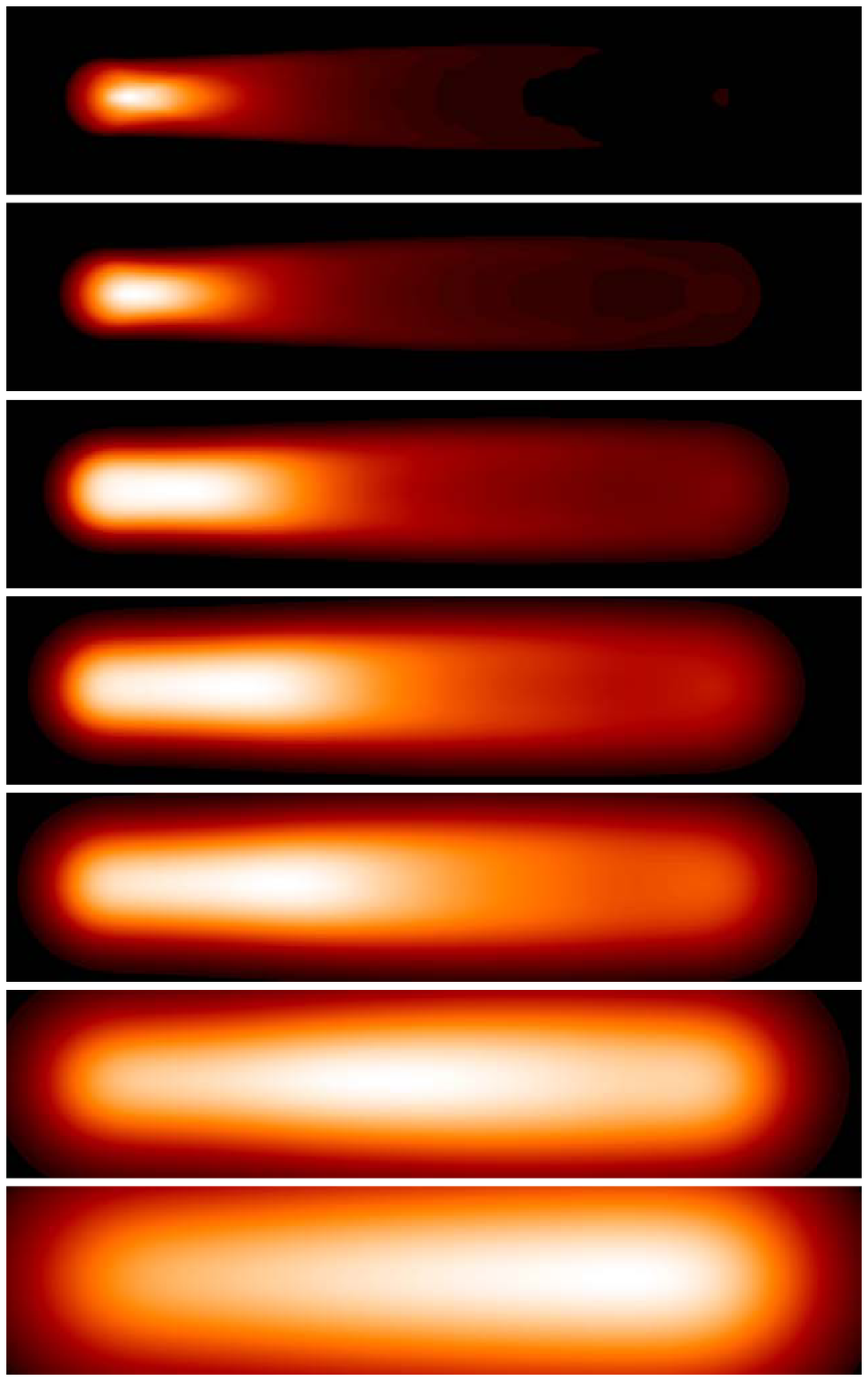}\hspace{0.2cm}\includegraphics[width=0.66\textwidth]{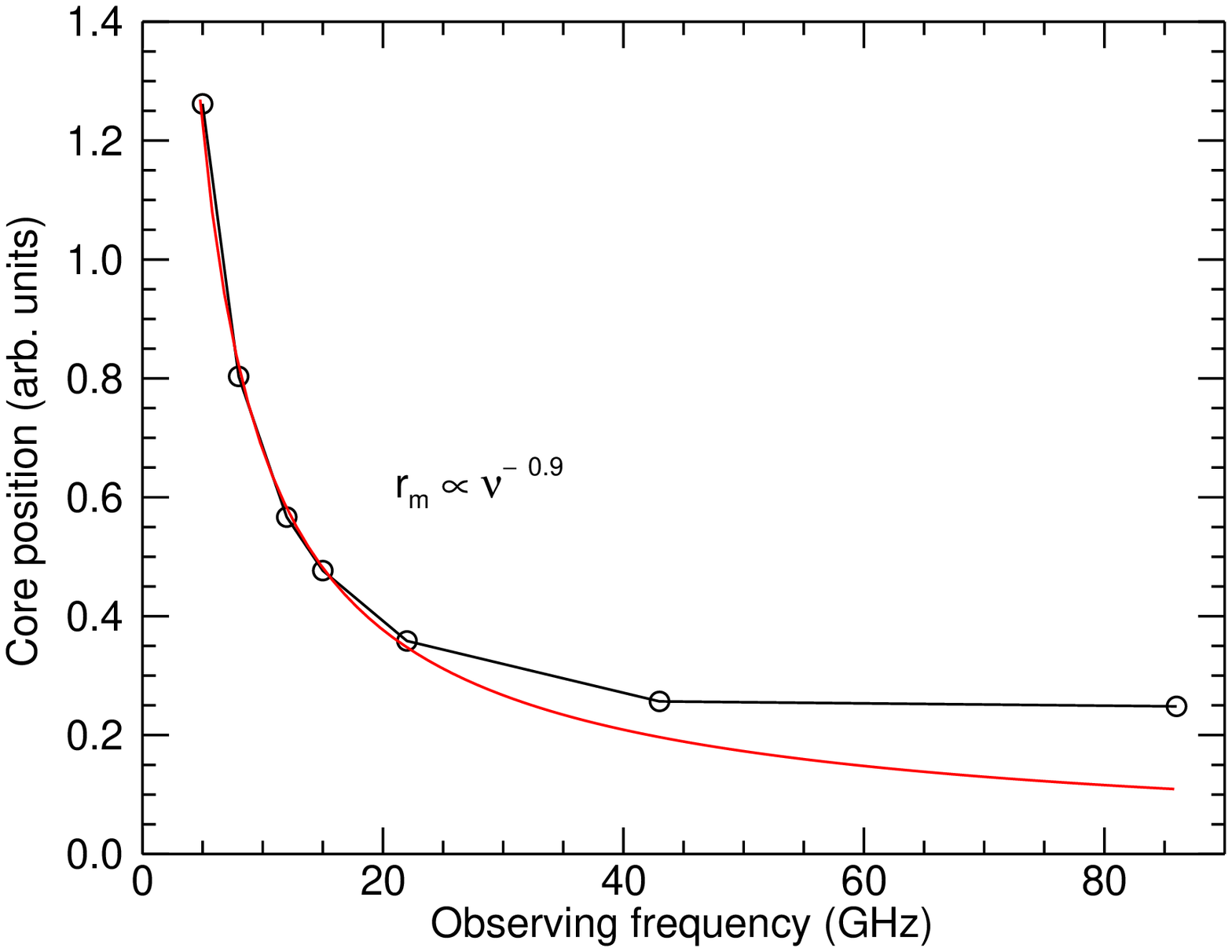}
\end{center}
\vspace{-0.2cm}
\caption{{\it Left:} Results from the numerical simulations. From top to bottom, a sequence of synchrotron total intensity images computed at 86, 43, 22, 15, 12, 8, and 5 GHz, respectively, using a relativistic hydrodynamical model of a jet with a recollimation shock. {\it Right:} Position of the core (peak emission) as a function of frequency. The red curve indicates the best fit to the core positions between 5 and 22 GHz, which follows the expected opacity core-shift of a conical Blandford \& K\"onigl jet model. The 43 and 86 GHz simulations clearly deviate from the opacity core-shift curve, revealing the fixed location of the recollimation shock associated with the core. 
\label{fig:cs_mod}}
\end{figure}

We have performed numerical simulations to test this proposed model, using the finite-volume code {\em Ratpenat}, which solves the equations of relativistic hydrodynamics \citep[][and references therein]{perucho_10}. The jet is launched with an initial over-pressure of 1.5 times that of the external medium in order to obtain a recollimation shock that can be identified with the core. Using the hydrodynamical results as input, we have then computed the synchrotron emission at different observing frequencies \citep[for details of the numerical model used see][]{gomez_95,gomez_97,2003ApJ...585L.109A}. This is illustrated by Fig.~\ref{fig:cs_mod}, which shows the sequence of total intensity images at the different frequencies, as well as the evolution of the core position with frequency. Full details of the simulations will be published elsewhere; here we just summarise the not-unexpected conclusions, which are that, at cm-wavelengths (5 to 22 GHz) the simulations reproduce the opacity core-shift of a Blandford \& K\"onigl conical jet model, while at mm-wavebands (43 and 86 GHz) the core position clearly departs from this behavior, revealing the recollimation shock at a fixed jet location.

Testing observationally whether the VLBI core at millimeter wavelengths is indeed consistent with the presence of a recollimation shock requires therefore bona-fide astrometric measurements of the core-shift spanning a wide range of centimeter and millimeter wavelengths. Confirmation of this model will support the hypothesis that the majority of the $\gamma$-ray flares in AGN jets are produced by the passing of new superluminal features through a pattern of recollimation shocks in the innermost jet regions, which would also include the VLBI core \citep{2013ApJ...773..147J,2015ApJ...813...51C,gomez_16}. If our model is correct, we expect to see the Blandford \& K\"onigl core-shift at cm-wavebands (5, 8.4, 15, and 22 GHz), with deviation from this behavior at mm-wavebands (43 and 86 GHz).
 { It should be also noted that at the innermost jet regions probed at mm-wavebands, other effects, such as radiative cooling or the parabolic jet shape found in M~87 by \citet{asada_12}, may also lead to a departure from the opacity core-shifts of the conical Blandford \& K\"onigl jet model. However none of these effects appear to affect the measurements performed by \cite{hada_11}, perhaps due to the progressive increase in the Doppler boosting as the jet accelerates in the innermost jet regions \citep{asada_14}.}
 { Additionally, recent space VLBI {\it RadioAstron} observations of BL~Lac at a record angular resolution of 21 $\mu$as have found evidence for the association of the radio core with a recollimation shock \citep{gomez_16}, providing extra motivation for this work.} 

Conventional phase referencing (PR) \citep{alef88,beasley} is the best approach for this analysis in cm-wavelength observations. The source/frequency phase referencing (SFPR) method works well for mm-wavelength observations, and has been demonstrated with frequencies as high as 130GHz (2mm) \citep{rioja_15}. However SFPR requires a second calibrator source with-in about 10$^o$ of the target. The density of calibrators at 86GHz, or even at 43GHz, is not sufficient to guarantee that a suitable source will be within this range. Indeed, this was the case for the source discussed in this paper: \bllac. 
Therefore we have developed a method built on the Frequency Phase Transfer (FPT) approach that under-pins SFPR, but does not require a second source. The description and validation of this method is the focus of this paper.

\section{Observations}\label{sec:obs}
The observations presented here were carried out on 2013 July 5 with the Very Long Baseline Array (VLBA) targeting the jet in BL Lac, and are part of a series of similar experiments on a sample of blazar sources aimed to test the correspondence of the mm-VLBI core with a recollimation shock. The analysis here provides the demonstration and explanation of a new technique we have developed for single source astrometric $\lambda$-astrometry mm-VLBI, where the high frequency images are astrometrically registered to a lower frequency image.

{ All observations were made at 2Gbps with 32MHz Intermediate Frequency (IF) bands}. Prime calibrators, 3C\,345 and 3C\,84, were observed for all frequency bands. Ten blocks of conventional phase referencing of \bllac\ at 5, 8, 15 and 22 GHz were performed, with \pcal\ and J2218+4146 as the reference sources and a cycle time of 80 seconds. Following each phase referencing block, we had fast frequency-switching observations, just on \bllac, between 22--43GHz and 22--86 GHz, with 30 seconds per scan over a 15 minute block. These were bracketed by ionospheric calibration blocks, which consist of observations switching between the L-band receiver range (16IFs between 1.4--1.7GHz), the wide band C-band receiver (16IFs between 3.9--7.9GHz) and the K-band receiver (16IFs between 21.8--22GHz), with 40 seconds of observing time at each band. { For the switching observations the on-source time was 1.3 hours for each frequency in the pair.}
{ The data reduction was in AIPS, following the standard path of correction for SEFD amplitudes, correction for Earth Orientation Parameters and correction for the Ionosphere, based on GPS models.}

\section{Multi-Frequency Phase Referencing}\label{sec:meth}
The Source Frequency Phase Referencing method, which has been described in detail elsewhere \citep{vlba_31,rioja_11a,rioja_11b,rioja_14,rioja_15,dodson_14}, consists of two calibration steps. In a first step, the observations at the higher frequency bands are calibrated using near-simultaneous \citep{vlba_31,rioja_11a,rioja_14} or simultaneous \citep{rioja_14,rioja_15,dodson_14} observations at a lower frequency band, for each source. This is done for all frequency pairs which have an integer\footnote{for non-integer ratios see the analysis and discussions in \citet{dodson_14}} frequency ratio, by which the low frequency calibration phase solutions are scaled. This dual frequency  calibration step eliminates the common non-dispersive residual errors (e.g. tropospheric propagation effects and geometric errors) in the complex visibility output of the correlator, providing an increased signal coherence at the higher frequency. 
The second step of the calibration removes the remaining dispersive residual errors (i.e. instrumental and ionospheric propagation effects) using the interleaved observations of another source. This two-step calibration  retains the astrometric signature of any source position shifts between the two frequencies in the interferometric phase observable.  
The Fourier transformation of the SFPR dataset is the SFPR map, which conveys a bona-fide astrometric measurement of the relative separation or shift between the 
reference points in the images at the two frequencies,  for the two sources. 
Results from SFPR analysis are to be found in the cited papers.

SFPR relies on the observation of a second \citep[or multiple, as in the analysis of][]{rioja_15} calibrator. This can be some distance from the target, with successful demonstrations with separations as large as 11\arcdeg \citep{vlba_31,rioja_15}, but there are sources for which one still struggles to find a suitable calibrator at the highest frequencies. This is the case for \bllac\ where no mm-wavelength calibrator, for direct or reverse SFPR, could be found within 10\arcdeg. Therefore we have attempted to achieve phase referencing in a similar fashion to SFPR, but without the second source. Our approach in this experiment is to calibrate all frequencies against a well-known source with precise astrometric position, then solve for the residual delays for the target, across a wide frequency span. This allows us to measure the residual TEC in the target direction, which is used to produce an ionosphere-free dataset for all frequencies.
This method we dub Multi-Frequency Phase Referencing (MFPR),
as now our calibration scheme allows relative astrometry between the (mm) frequency bands corrected by observations at multiple (cm) frequencies of the target. 

\section{Methods}\label{sec:method}
The prime calibration was against 3C\,345, except for 86 GHz, where we had many missed scans and the data quality was very poor. For this frequency 3C\,84 was the prime calibrator and { prime calibration} could only be performed for the BR, KP, LA, OV and PT antennas. Prime calibration removes the instrumental terms, plus all the atmospheric contributions in the direction of the calibrator, at the time of the observations. We handled the structural contribution from the source by hybrid mapping the data before astrometric calibration, to produce a reference image which is used in the analysis. The conventional phase referencing at cm-wavebands was calibrated following standard procedures
and the detailed interpretation will be reported elsewhere \citep{molina_bllac}. 

For the ionospheric correction blocks we use the delay (only) from each IF to measure the Total Electron Content (TEC) contribution on the line of sight towards \bllac, as a function of time. { This is a measurement of the residual ionospheric contribution, after correction with the GPS data and the subtraction of the TEC in the direction of the prime calibrator, at the time of that scan.}
We fitted a linear slope in { $\nu^{-2}$} to the semi-simultaneous delay measurements (i.e. such that $\tau(\nu)=\tau_{\rm trop}+\tau_{\rm iono} \nu_{GHz}^{-2}$) and calculate the residual TEC contribution as a function of time, for that line of sight, using $\Delta$TEC=0.75\,$\tau_{\rm iono}$. Here $\tau(\nu)$ are the measured delays as a function of frequency for one block of ionospheric calibration observations, $\tau_{\rm trop}$ is the non-dispersive delay { (from both clock and tropospheric contributions)}, $\tau_{\rm iono}$ is the ionospheric delay (at 1 GHz), $\nu_{GHz}$ is the frequency in GHz and $\Delta$TEC is the deduced residual ionospheric contribution in Total Electron Count Units (TECU), to the line of sight of the target. { We are assuming, as we have corrected the instrumental terms ($\tau_{\rm inst}$) using the prime calibration, that these are constant during the experiment and therefore can be ignored in this analysis.} The derived $\Delta$TEC is then used to calculate the ionospheric contribution for each IF, at each time interval. The frequency-dependent delay can directly be calculated, but note that the sign of the phase has to be reversed, as the ionospheric contribution is a group delay not a phase delay. 
Finally we solved for the (ionosphere-free) delay, rate and phase on the (ionosphere-corrected) 22 GHz data and applied these solutions, suitably scaled by the frequency ratio, to the ionosphere-corrected 43 and 86 GHz data. 
{ As this calibration scheme uses low frequency calibration blocks, either side of the high frequency science observations, and these are to correct for the ionospheric contributions to the atmosphere, we dubbed these IonospheriC Excision
blocks (ICE-blocks). }
This recognises the commonality with the so called Geodetic Block calibration schemes that dealt with the static tropospheric contributions \citep{brunthaler_05,honma_08, reid_micro}. 

Our expectations were that, having corrected for the ionospheric and the tropospheric contributions, we should be left with the high frequency datasets astrometrically aligned to the 22 GHz dataset.
However, as will be discussed in the results, we found that our initial calibration scheme was inadequate. In our observations of 3C\,345 we have a gap of about five minutes between the prime calibration observations at 22 and 43 GHz, which was used for a pointing correction. An even longer gap existed between the 22 and 86 GHz prime-calibrator scans. Therefore we could not align the phase of these scans to a common point in time. This caused an unknown phase offset to be introduced between the data at 22 GHz and higher frequencies, in the observation of the target source at the target frequencies. Therefore prime-calibration should have included a fast frequency-switching scan on the calibrator as well as the target; this will be included in future observations. { How we resolved this problem for these observations is discussed in Section \ref{sec:res:off}.}

\section{Results}\label{sec:res}
\subsection{Ionospheric measurements}
\begin{figure}[ht]
\centering
\includegraphics[width=10cm]{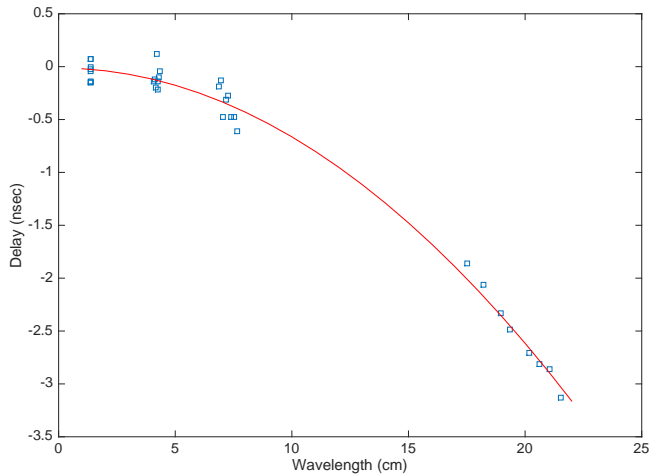}
\caption{A typical delay $\tau(\nu)$ for one particular antenna (BR) and 2 minute ICE-block scan (UT 05:56), as a function of wavelength. It shows the curvature (following $\nu^{-2}$) that directly measures the $\Delta$TEC residual (in this case -4.4 TECU, indicated with the solid line), for that solution interval and for that antenna, in the line of sight of the target. 
\label{fig:tec_mu}}
\end{figure}

\begin{figure}[ht]
\centering
\includegraphics[width=10cm]{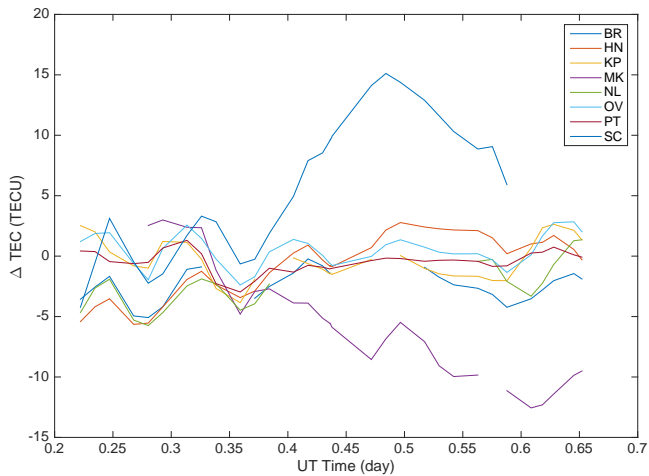}
\caption{The $\Delta$TEC residuals for all antennas across the duration of the experiment. The values for most antennas range between $\pm$5\,TECU, as would be expected for data that has been corrected with the default TEC maps, using TECOR. The antennas MK and SC show the largest deviations. Errors in individual measurements are typically 0.1\,TECU. 
\label{fig:tec_t}}
\end{figure}

Each ICE-Block consists of multiband observations at 1.4--1.7, 3.9--7.9 and 22 GHz, that is 21 to 18 cm, 8 to 4 cm, and 1.3 cm. These bracket the fast frequency-switching observations of the target source, which either consist of 22--43 or 22--86 GHz blocks. The ICE-blocks are calibrated following standard procedures and the measured delays for each IF fitted to derive $\Delta$TEC. Figure \ref{fig:tec_mu} shows the delays as a function of frequency for one scan on one antenna. Figure \ref{fig:tec_t} shows the derived $\Delta$TEC for all antennas during this experiment. { The fitting errors can be used to estimate the measurement precision of the $\Delta$TEC residuals. The error level was $\sim$0.1\,TECU, if one excluded MK and SC (which were not included in the final MFPR imaging, due to the poor quality of the data) and 0.2\,TECU if all antennas are included. 0.1\,TECU would contribute 2.2$^\circ$ of phase at 22 GHz, which would be scaled up to 3.3$^\circ$ and 8.2$^\circ$ of phase noise at 43 and 86 GHz, respectively \citep{rioja_11a}.}


\subsection{Constant Phase Offset measurement and Astrometric Results}\label{sec:res:off}
\begin{figure}[hbt]
\centering
\includegraphics[width=8cm]{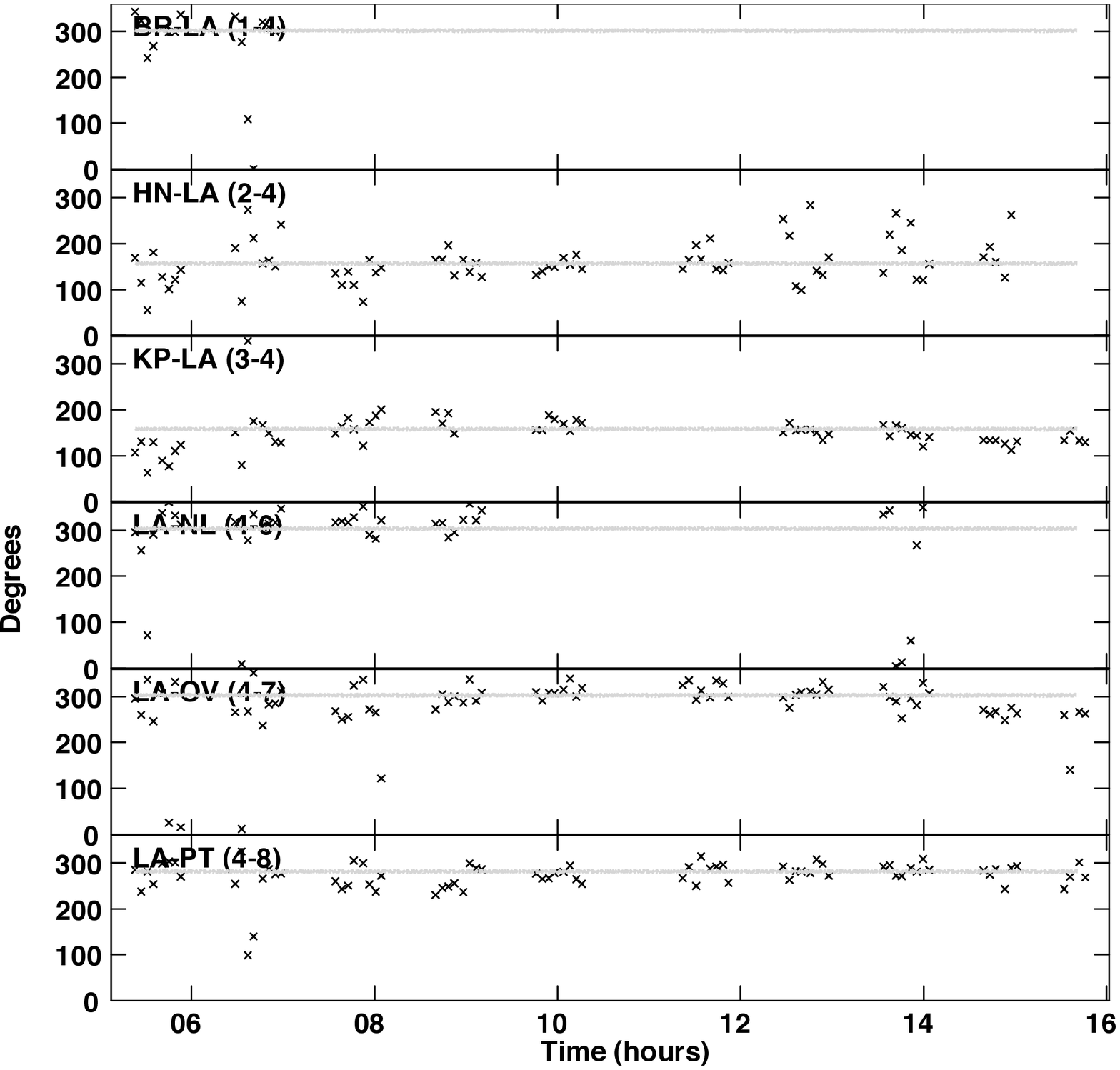}
\includegraphics[width=8cm]{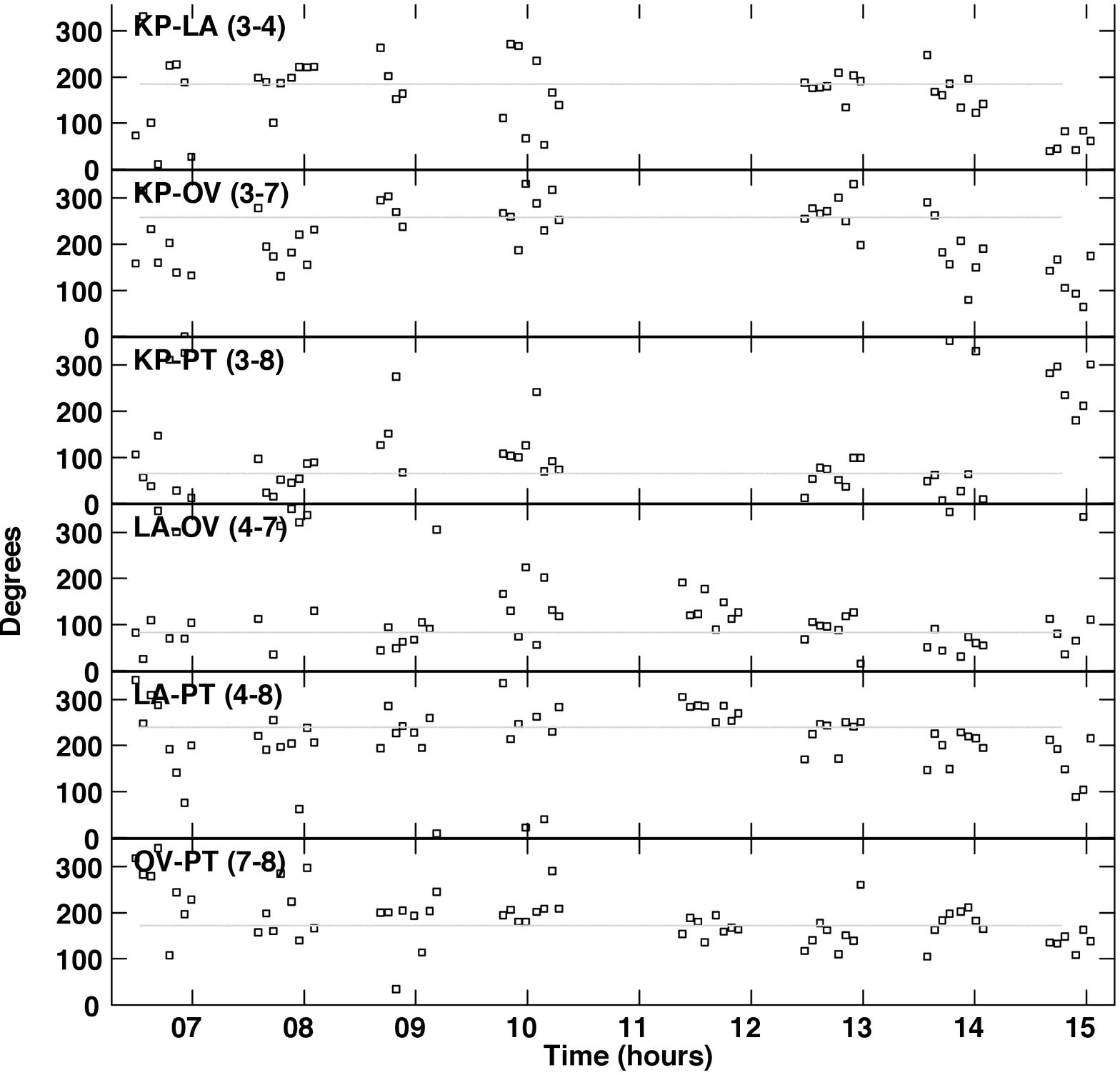}
\caption{Ionosphere and Troposphere corrected visibilities on \bllac\ showing the dominant constant phase offset (indicated with light grey lines) arising from the non-simultaneous observation of the prime calibrator, for {\em left} 43 GHz with baselines to Los Alamos (LA) only and {\em right} 86 GHz with all baselines that could be calibrated.
\label{fig:fpt}}
\end{figure}

After subtraction of the calculated ionospheric delays and phases from the 22, 43, and 86 GHz datasets, we self-calibrated the 22 GHz \bllac\ data against a hybrid map of the source, scaled the solutions by the frequency ratio and use these to correct the higher frequencies. Thus the ionospheric and tropospheric contributions are removed. However, because the initial prime-calibrations at different frequencies were not at close points in time, there is an introduced constant, but unknown, phase at each station. This is shown in Fig.~\ref{fig:fpt} for 43 and 86 GHz.
The Fourier inversion of this data, after excluding antennas MK and SC plus the low elevation data (the first and last hour), gives us our initial image. However this image did not recover the source structure, because of the error in the initial calibration chain.

\begin{figure}[hbt]
\centering
\includegraphics[width=8cm]{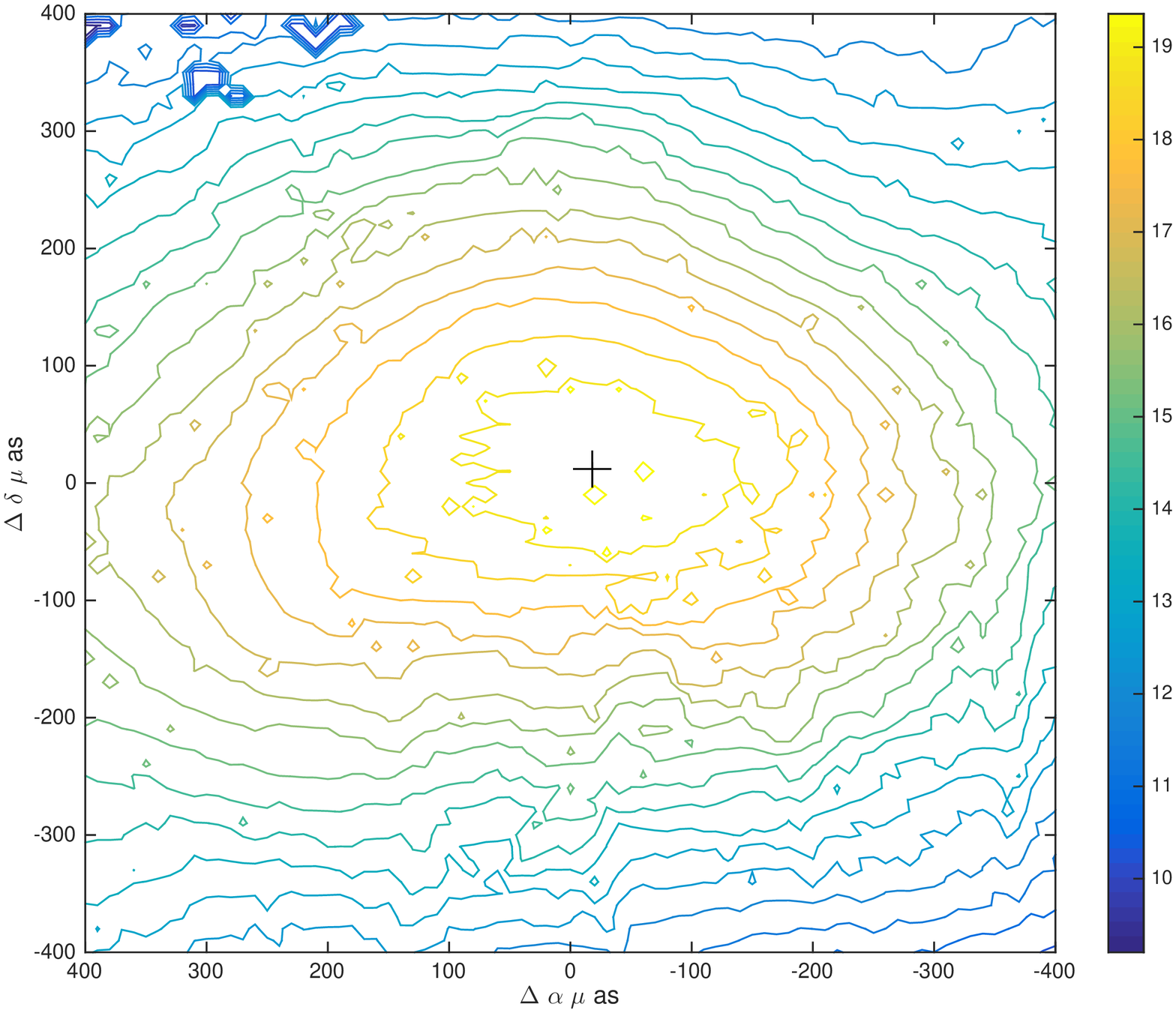}
\caption{The dynamic range as a function of the $\Delta \alpha$,$\Delta \delta$ of the input model. Also marked is the peak value, at -18, 12 $\mu$as, and the error bounds of $\pm$10 $\mu$as.
\label{fig:quality_dr2}}
\end{figure}

\begin{figure}[hbt]
\centering
\includegraphics[width=8cm]{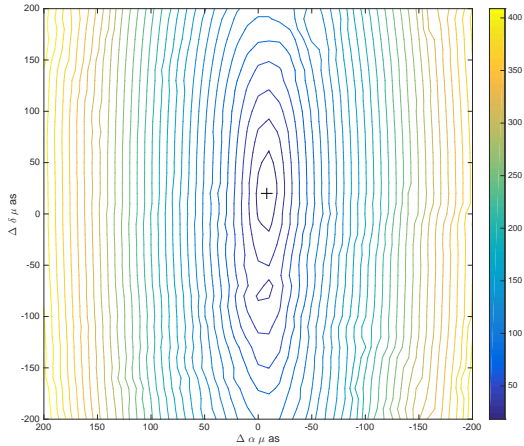}
\caption{The alignment of the requested model and the resultant peak of emission, as a function of the $\Delta \alpha$,$\Delta \delta$ of the input model. Also marked is the best alignment at -8, 20 $\mu$as and the error bounds of $\pm$5, 6 $\mu$as.
\label{fig:quality_align}}
\end{figure}

\begin{figure}[hbt]
\centering
\includegraphics[width=8cm]{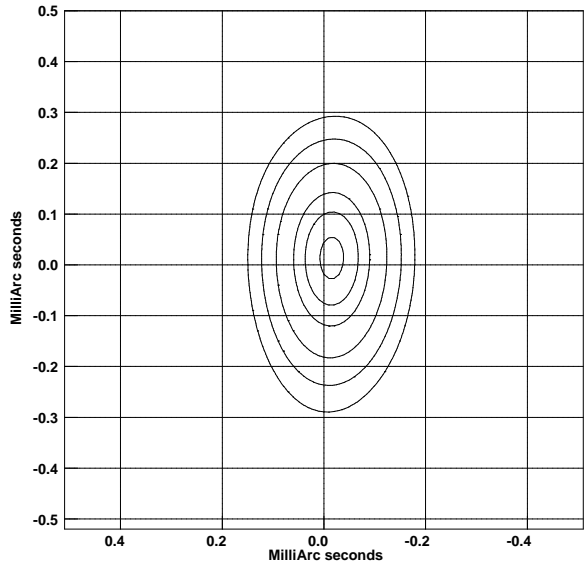}
\caption{Image of \bllac\ at 43 GHz, Multi-frequency Phase Referenced to the 22 GHz data. The offset from centre (-8,+20 $\mu$as) is the bona-fide astrometric core shift between 22 and 43 GHz. Contours are 60, 70, 80, 95 and 99\% of the peak flux (3.3 Jy/beam). The restoring beam size is 0.25$\times$0.34, PA -22$^o$.
\label{fig:qband_mfpr}}
\end{figure}

To remove a constant phase yet not lose the astrometric signal we trialed all possible models (point sources on a 10 $\mu$as grid within 0.5 mas of the phase centre), generating a single phase correction for each antenna for the whole dataset, using self-calibration. Using this calibration table we repeated the imaging and inspected the peak flux, the residual RMS, and the dynamic range as a function of the model. The peak in the dynamic range, plotted in Fig.~\ref{fig:quality_dr2}, gives an astrometric alignment of -18$\pm$10,+12$\pm$10 $\mu$as, based on the 95\% confidence limits of a 2D second order polynomimal fit to the surface.
{ Note that the dynamic ranges achieved (a maximum of $\sim$20) for a single point source fit are much less than those for the self-calibrated hybrid images, for which we had a dynamic range of $\sim$200.}
The most precise test, however, is the alignment of the location of the peak flux and that of the trial model. The minimum absolute offset in the alignment gives an astrometric result of -8$\pm$5,20$\pm$6 $\mu$as, based on the 95\% confidence limits of two orthogonal 1D second order polynomial fits to the surface, as shown in Figure \ref{fig:quality_align}. The precision of the alignment is approximately double that of the precision from the peak in the dynamic range
so we adopt this value as our astrometric result. The image of \bllac\ made with the phase corrections for an offset of -8$\pm$5,20$\pm$6 $\mu$as is shown in Fig.~\ref{fig:qband_mfpr}. 
{ Unfortunately the 86 GHz data was of too poor quality to produce useful results}. In a subsequent paper we will extend our analysis to 86 GHz using a better dataset from our sample of blazars included in this project.


\begin{figure}[hbt]
\centering
\includegraphics[width=8cm]{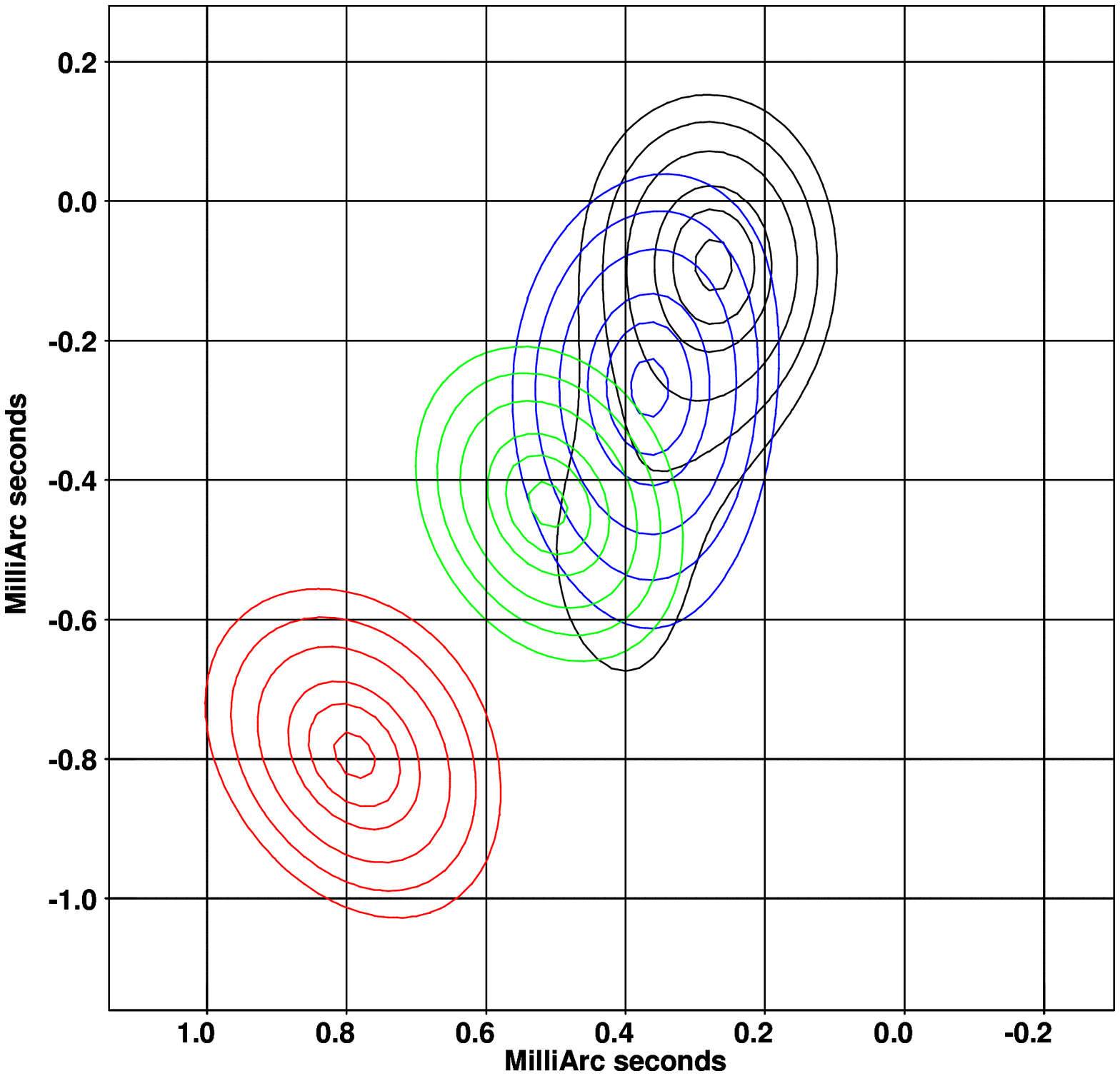}
\caption{Overlaid astrometrical-registered images of \bllac\ for the nominal phase centre of 22:02:43.291 and 42:16:39.980 in RA and Declination, derived using conventional methods and phase referenced to \pcal. Four frequencies are plotted: 4.8, 8.4, 15.2 and 21.9 GHz in red, green, blue and black respectively. Contours are 60, 70, 80, 95 and 99\% of the peak flux for each frequency, which are 1.6, 3.7, 2.1 and 1.6 Jy/beam.
\label{fig:conv_pr}}
\end{figure}

\begin{figure}[hbt]
\centering
\includegraphics[width=8cm]{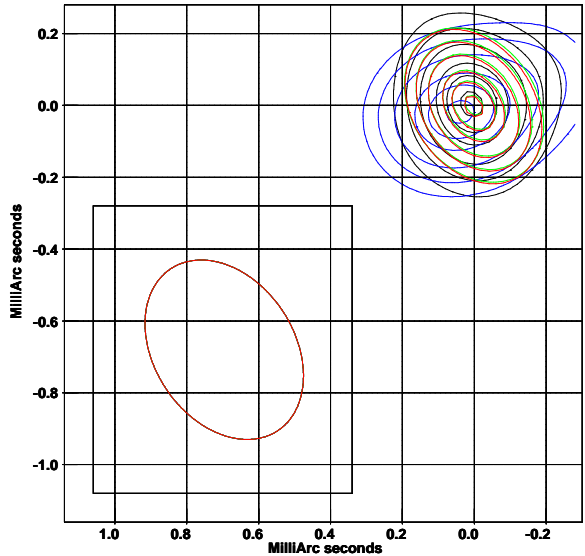}
\caption{Overlaid self-calibrated images of \pcal\ for the nominal phase centre of 21:53:50.959 and 43:22:54.500 in RA and Declination. All images are restored with the 22 GHz beam parameters of the image (0.57$\times$0.72, PA -26$^o$) used to reference the 43 GHz data. Four frequencies are plotted: 4.8, 8.4, 15.2 and 21.9 GHz in red, green, blue and black respectively. Contours are 60, 70, 80, 95 and 99\% of the peak flux for each frequency, which are 0.25, 0.28, 0.12 and 0.08 Jy/beam.
\label{fig:conv_pr3}}
\end{figure}

\begin{figure}[ht]
\centering
\includegraphics[width=8cm]{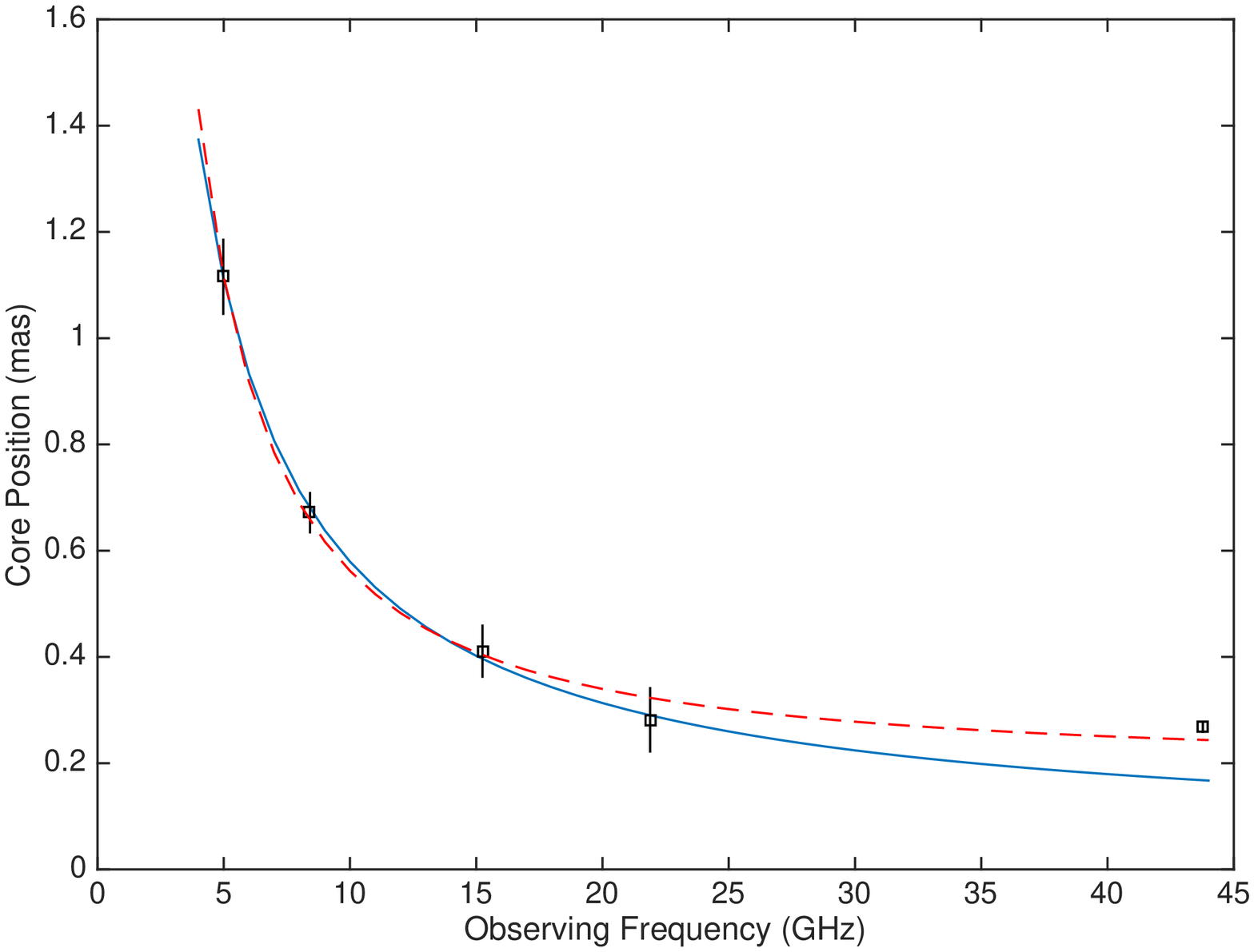}
\caption{Astrometric core-shifts of \bllac, derived using conventional phase referencing between 4.8 and 22 GHz, plotted as a function of frequency. All frequencies were restored with the beamsize at 22 GHz. Additionally we plot the position of the 43 GHz peak emission, derived using our MFPR method between 22/43GHz, and added to the position of peak emission of the 22 GHz. Error bars show estimates for the accuracy in conventional phase referencing (beam-width over dynamic range), except for the MFPR result, where the errors are from the measured precision given in the text.
Overlaid in blue is the model from the fitting of the cm-wavelength data (where $\kappa$ is -0.99 and $r_0$ is 5.3 mas\,GHz$^{\kappa}$) and in red dashes the fit for all data (in which case $\kappa$ is -1.32 and $r_0$ is 7.7 mas\,GHz$^{\kappa}$) 
\label{fig:mm2cm_cs}}
\end{figure}

To test the physicality of our results we compared the derived 22--43 GHz core-shift measurement to the preliminary results from the cm-wavelength phase referencing. We performed conventional phase referencing for the 4.8, 8.4, 15, and 22 GHz (see Fig.~\ref{fig:conv_pr}) data against the compact calibrator \pcal\ (Fig.~\ref{fig:conv_pr3}), which is 2$^\circ$ from \bllac. The images are similar and in agreement with published observations. We restored the \bllac\ images to the common beamsize of the 22 GHz image used as the reference for the FPT to avoid blending issues that otherwise shift the apparent positions.
We measured the relative positions of the peak of emission (compared to that of the referenced source) at all frequencies and found a systematic shift across the sky between 4.8 and 22 GHz, with $r_0$ of 5.3 mas\,GHz$^{\kappa}$, $\kappa$ of -0.99 and an offset of 45 $\mu$as. These parameters would lead to a prediction of 120$\pm$30 $\mu$as for the core-shift between 22 and 43 GHz. This is much greater than found.
Additionally we fitted for all frequencies, including 43 GHz. The best fit was obtained with $r_0$ of 7.7 mas\,GHz$^{\kappa}$, $\kappa$ of -1.32 and an offset of 190 $\mu$as. This fit, although valid, { would imply a significant departure from the expected value of $\kappa \simeq$-1, for the case of equipartition between jet particle and magnetic field energy densities \citep[e.g.,][]{lob_98}, as found at cm-wavebands in BL~Lac and other multiple sources \citep[i.e.,][]{osullivan_09,sokolovsky_11,hada_11}. It would also imply a large} error in the astrometric position of \bllac\ and/or \pcal, which is unlikely.
Figure \ref{fig:mm2cm_cs} shows the location of the peaks of emission, along with predictions from the two models.


\section{Discussions}\label{sec:disc}

{ We have demonstrated that the effects of the atmosphere can be calibrated in a step-wise manner, by decomposing them into dispersive and non-dispersive contributions.}
We have included for the first time a wide-band measurement of the residual ionospheric contributions to VLBI data, for the line of sight of the target. 
Our measurement of the ionosphere, derived from the group delay curvature, is a development of the standard geodetic two point fit (at 2.4 and 8.4-GHz) approach \citep{sxfeeds,new_geodetic}, whereas we fit the data over multiple frequencies. 
We performed simple simulations with typical values for $\Delta$TEC and $\Delta\ell$, the residual TEC and path length, and measurement errors of 0.1\,nsec (1$^\circ$ phase error across 16 MHz). We explored the required frequency span to be able to predict accurate values for $\Delta$TEC, and found that the lower frequencies are the most crucial. At least 2.4 GHz would be required, and 1.4 GHz would be preferred. 
{ Using our data we investigated} the achievable { reliability} using just the new VLBA wideband C-band and K-band data, and found the accuracy in the determined $\Delta$TEC to be $\sim$1\,TECU. { Such levels are those expected (i.e. by scaling with $(\nu_1/\nu_2)^2$) and would  be sufficient mitigate the ionospheric contribution to the errors in the astrometry of methanol masers at 6.7\,GHz.}

The Frequency Phase Transfer has been thoroughly demonstrated previously \citep [and references therein]{rioja_15}, 
therefore we would expect the approaches taken here to be successful. The conditions for which we may expect this to break down are: where instrumental terms are not stable, introducing a time variable non-dispersive term that is not a function of { $\nu^{-2}$}; and where there are baseline dependent terms introduced, by for example having a poor model of the prime calibrator. 
For our analysis we do not believe the first of these are issues, as the VLBA has extremely stable instrumental terms. 

{ We addressed the issue of the baseline dependent phase terms introduced by the prime-calibrator by using a hybrid model in the initial fringe fitting stage.} This worked well for the core-dominated source 3C\,345, but failed for the more complex structure of 3C\,84, where there were insufficient data at 86 GHz to constrain the model.
We note that any core-shift in the prime calibrator will appear as a constant phase introduced between the two frequencies, as it will arise from a single scan. { Therefore it will be absorbed into the constant phase corrections discussed next.} 

To remove the phase introduced between the two frequencies from the prime-calibration we performed a grid search to find the best constant phase that matched the data.
Our approach of testing for model stability is similar to super resolution. 
In super-resolution one adjusts the model to the uv-data to locate the best fitting location using a minimisation method, thereby producing a positional accuracy greater than the resolution. 
In our case we are testing the fit of a model to the uv-data, with a cycle of self-calibration, in a stepwise fashion, over the parameter space.
{\bf This aspect of our analysis is perhaps the most innovative, but also suffers from the limitations in our minimisation methods. We have no mechanism to measure the co-variance between the fitted parameters of source position and constant station-based phase offset. If the experiment had been a `snapshot' these two would be degenerate, however as the observations spanned 8 hours this degeneracy is broken. We estimated our errors from the fitting to the 2D surface of the results. The use of independent measures for these surfaces reduces the possibility of degeneracy. That the maximum dynamic range (and peak flux and minimum residual RMS) align with those of the minimum absolute offset in the alignment gives confidence in our results.}

Our conclusions would be significantly strengthened if we were able to produce a joint analysis of the 22 to 43 and 86 GHz data. However this was not possible, because of the poor quality of the observations at 86 GHz, due to technical issues. These data could not even be self-calibrated to produce acceptable images.
Therefore we believe the residuals probably arise from baseline-based (rather than station based) contamination from the prime-calibrator scan, which can not be corrected for by our procedures. 

Ideally we would compare our new method with results from conventional phase referencing for the same source. However if conventional phase referencing was possible at these frequencies we would not have needed to develop these new methods. 
{ We used the measurement of the core-shifts made at lower frequencies with conventional phase referencing, and extrapolated these results to compare with our MFPR measurement made at mm-wavelengths.} However these can provide only the Blandford \& K\"onigl core-shift; our expectation is that the higher frequencies would have smaller than predicted core-shifts. This is consistent with what we discover, as shown in Fig.~\ref{fig:mm2cm_cs}. Consequently we can not use this approach to validate our method.

We attempted to derive the core shift following the approach of aligning optically thin features, but we could not get sufficiently accurate results from our own data. { Optically thin features tend to be of lower surface brightness and therefore it is difficult to accurately determine the centroids  \citep{hovatta_14}.} Nevertheless cross-correlation of VLBA images firstly by \citet{osullivan_09}, and secondly by \citet{gomez_16} (with observations made a few months after ours in November 2013) have determined a core-shift of 30$\pm$20 $\mu$as 
and 21 $\mu$as, respectively, between 22 and 43 GHz. 
This is in close agreement with the value we find from the MFPR analysis, which provides extra support for the reliability of our new method.
The alignment reported by \citet{osullivan_09} is based on the cross correlation of images at frequencies between 5 and 43 GHz. They find an average core-shift of the 22GHz position, referenced to 43 GHz, of 40$\pm$20$\mu$as, whereas we have compared with only the measurement between 22-43 GHz. All measurements are consistent within the errors. They fitted their cross-correlation alignments to obtain a value of $\kappa$ (following our definition) of -1.01 between 5 and 43GHz, which would a first sight appear to be in contradiction with our results. However the functional expression ($(\nu_2-\nu_1)/(\nu_1 \nu_2)$) that they fit is not very sensitive to deviations between 22 and 43 GHz. { Replacing all of their 43GHz measurements with the 22GHz values (i.e. inserting a zero core-shift between 22 and 43GHz) and repeating the fitting reproduces the published result, within errors.} Furthermore, in our analysis, we found that blending of components within the uniform weighted beam at the low frequencies distorted our results, which is why we restored with the super resolved beam of the 22 GHz. This important issue was not addressed in the \citet{osullivan_09} analysis. 

{ This multi-frequency approach is one of two methods we are currently developing to improve astrometric calibration of the ionospheric contributions, the other being to use multiple calibrators around the target and to solve for the spatial structure of the atmosphere, which we call {\em MultiView} \citep{rioja_09,rioja_16}. This method, which requires more calibrators rather than less, simultaneously solves for a 2D Ionospheric and Tropospheric phase screen over the array, and will be extremely suitable for VLBI stations with multiple beams, such as SKA and ASKAP.}

\section{Conclusions}\label{sec:conc}

We have presented a development of the SFPR method, which we call Multi-Frequency Phase Referencing (MFPR). This has been used to measure the core-shift (or the $\lambda$-astrometry) for \bllac, between 22 and 43 GHz. 
The MFPR method presented here involved a measurement of the $\Delta$TEC on the line of sight of the target, with precision of 0.1\,TECU. Once the data is corrected for the ionosphere the tropospheric correction is measured at the lower frequency and applied to the higher frequency.  
The high frequency data should then be registered to the low frequency data using Frequency Phase Transfer.
Our data required an additional step, to minimise over a constant phase offset introduced by the prime calibration, which can be avoided in future observations with improved scheduling.
After these steps the astrometric offset between the two frequencies is the bona-fide core-shift, which has been derived from observations of a single source by careful calibration of the atmospheric contributions.

This new method opens up a large number of possibilities for astrometric analysis in mm-VLBI. Conventional phase referencing is not possible for mm-VLBI, and for a significant number of sources, such as \bllac, a suitable calibrator as required for SFPR can not be found. The method of alignment of optically thin components { suffers from both questionable validity, and a requirement of high sensitivity} to detect low surface brightness features.  MFPR bypasses all of these issues, requiring only a detectable target source at the lower frequencies. With simultaneous observations this method will be applicable to mm and sub-mm wavelengths.

The results from the MFPR measures a core-shift for \bllac\ of -8$\pm$5,20$\pm$6 $\mu$as between 22 and 43 GHz. This is significantly less than the prediction from measurements of the core-shift at cm-wavelengths, but in line with both the theoretical expectations and other work.

Further analysis, to be published in \citet{molina_bllac}, will improve the initial measurement of the core-shift at cm-wavelengths, allowing us to deduce if we are truly uncovering deviations from the Blandford \& K\"onigl model. 
If so this would be the first detection of the predicted association of the mm-VLBI core with a recollimation shock responsible for the $\gamma$-ray emission in blazar { jets, in agreement with the findings of \cite{gomez_16}.}
The fact that the calibration of the 86 GHz showed promise but was defeated by an unusually large antenna failure rate gives us confidence that we will be able to perform MFPR at 86 GHz in future demonstrations.

\noindent
{\bf Acknowledgements}

\noindent
We would like to especially acknowledge the numerous conversations with Richard Porcas about the best approaches for the analysis, and the referee for comments that improved the paper. 
The work at IAA-CSIC is supported by the Spanish Ministerio de Econom\'{\i}a y Competitividad grant AYA2013-40825-P.
The VLBA is operated by National Radio Astronomy Observatory and is a facility of the National Science Foundation operated under cooperative agreement with Associated Universities Inc. 
\pagebreak


\begin{thebibliography}{51}
\expandafter\ifx\csname natexlab\endcsname\relax\def\natexlab#1{#1}\fi

\bibitem[{{Alef}(1988)}]{alef88}
{Alef}, W. 1988, in IAU Symposium, Vol. 129, The Impact of VLBI on Astrophysics
  and Geophysics, ed. M.~J. {Reid} \& J.~M. {Moran}, 523

\bibitem[{{Aloy} {et~al.}(2003){Aloy}, {Mart{\'{\i}}}, {G{\'o}mez}, {Agudo},
  {M{\"u}ller}, \& {Ib{\'a}{\~n}ez}}]{2003ApJ...585L.109A}
{Aloy}, M.-{\'A}., {Mart{\'{\i}}}, J.-M., {G{\'o}mez}, J.-L., {et~al.} 2003,
  \apjl, 585, L109

\bibitem[{{Asada} \& {Nakamura}(2012)}]{asada_12}
{Asada}, K., \& {Nakamura}, M. 2012, \apjl, 745, L28

\bibitem[{{Asada} {et~al.}(2014){Asada}, {Nakamura}, {Doi}, {Nagai}, \&
  {Inoue}}]{asada_14}
{Asada}, K., {Nakamura}, M., {Doi}, A., {Nagai}, H., \& {Inoue}, M. 2014,
  \apjl, 781, L2

\bibitem[{{Beasley} \& {Conway}(1995)}]{beasley}
{Beasley}, A.~J., \& {Conway}, J.~E. 1995, in Astronomical Society of the
  Pacific Conference Series, Vol.~82, Very Long Baseline Interferometry and the
  VLBA, ed. J.~A. {Zensus}, P.~J. {Diamond}, \& P.~J. {Napier}, 327

\bibitem[{{Blandford} \& {K{\"o}nigl}(1979)}]{bk_79}
{Blandford}, R.~D., \& {K{\"o}nigl}, A. 1979, \apj, 232, 34

\bibitem[{{Brunthaler} {et~al.}(2005){Brunthaler}, {Reid}, {Falcke},
  {Greenhill}, \& {Henkel}}]{brunthaler_05}
{Brunthaler}, A., {Reid}, M.~J., {Falcke}, H., {Greenhill}, L.~J., \& {Henkel},
  C. 2005, Science, 307, 1440

\bibitem[{{Casadio} {et~al.}(2015){Casadio}, {G{\'o}mez}, {Jorstad},
  {Marscher}, {Larionov}, {Smith}, {Gurwell}, {L{\"a}hteenm{\"a}ki}, {Agudo},
  {Molina}, {Bala}, {Joshi}, {Taylor}, {Williamson}, {Arkharov}, {Blinov},
  {Borman}, {Di Paola}, {Grishina}, {Hagen-Thorn}, {Itoh}, {Kopatskaya},
  {Larionova}, {Larionova}, {Morozova}, {Rastorgueva-Foi}, {Sergeev},
  {Tornikoski}, {Troitsky}, {Thum}, \& {Wiesemeyer}}]{2015ApJ...813...51C}
{Casadio}, C., {G{\'o}mez}, J.~L., {Jorstad}, S.~G., {et~al.} 2015, \apj, 813,
  51

\bibitem[{Casadio {et~al.}(2015)Casadio, G{\'o}mez, Grandi, Jorstad, Marscher,
  Lister, Kovalev, Savolainen, \& Pushkarev}]{2015ApJ...808..162C}
Casadio, C., G{\'o}mez, J.~L., Grandi, P., {et~al.} 2015, ApJ, 808, 162

\bibitem[{{Chatterjee} {et~al.}(2011){Chatterjee}, {Marscher}, {Jorstad},
  {Markowitz}, {Rivers}, {Rothschild}, {McHardy}, {Aller}, {Aller},
  {L{\"a}hteenm{\"a}ki}, {Tornikoski}, {Harrison}, {Agudo}, {G{\'o}mez},
  {Taylor}, \& {Gurwell}}]{chatterjee_11}
{Chatterjee}, R., {Marscher}, A.~P., {Jorstad}, S.~G., {et~al.} 2011, \apj,
  734, 43

\bibitem[{{Daly} \& {Marscher}(1988)}]{daly_88}
{Daly}, R.~A., \& {Marscher}, A.~P. 1988, \apj, 334, 539

\bibitem[{{Dodson} \& {Rioja}(2009)}]{vlba_31}
{Dodson}, R., \& {Rioja}, M.~J. 2009, {VLBA Scientific Memorandum n. 31:
  Astrometric calibration of mm-VLBI using ''Source/Frequency Phase
  Referenced'' observations}, Tech. rep., NRAO

\bibitem[{{Dodson} {et~al.}(2014){Dodson}, {Rioja}, {Jung}, {Sohn}, {Byun},
  {Cho}, {Lee}, {Kim}, {Kim}, {Oh}, {Han}, {Je}, {Chung}, {Wi}, {Kang}, {Lee},
  {Chung}, {Kim}, {Kim}, {Lee}, {Roh}, {Oh}, {Yeom}, {Song}, \&
  {Kang}}]{dodson_14}
{Dodson}, R., {Rioja}, M.~J., {Jung}, T.-H., {et~al.} 2014, \aj, 148, 97

\bibitem[{{Fromm} {et~al.}(2015){Fromm}, {Perucho}, {Ros}, {Savolainen}, \&
  {Zensus}}]{fromm_15}
{Fromm}, C.~M., {Perucho}, M., {Ros}, E., {Savolainen}, T., \& {Zensus}, J.~A.
  2015, \aap, 576, A43

\bibitem[{{G{\'o}mez} {et~al.}(1997){G{\'o}mez}, {Mart{\'{\i}}}, {Marscher},
  {Ib{\'a}{\~n}ez}, \& {Alberdi}}]{gomez_97}
{G{\'o}mez}, J.~L., {Mart{\'{\i}}}, J.~M., {Marscher}, A.~P., {Ib{\'a}{\~n}ez},
  J.~M., \& {Alberdi}, A. 1997, \apjl, 482, L33

\bibitem[{{G{\'o}mez} {et~al.}(1995){G{\'o}mez}, {Marti}, {Marscher}, {Ibanez},
  \& {Marcaide}}]{gomez_95}
{G{\'o}mez}, J.~L., {Marti}, J.~M.~A., {Marscher}, A.~P., {Ibanez}, J.~M.~A.,
  \& {Marcaide}, J.~M. 1995, \apjl, 449, L19

\bibitem[{{G{\'o}mez} {et~al.}(2016){G{\'o}mez}, {Lobanov}, {Bruni}, {Kovalev},
  {Marscher}, {Jorstad}, {Mizuno}, {Bach}, {Sokolovsky}, {Anderson}, {Galindo},
  {Kardashev}, \& {Lisakov}}]{gomez_16}
{G{\'o}mez}, J.~L., {Lobanov}, A.~P., {Bruni}, G., {et~al.} 2016, \apj, 817, 96

\bibitem[{{Guirado} {et~al.}(1995){Guirado}, {Marcaide}, {Alberdi}, {Elosegui},
  {Ratner}, {Shapiro}, {Kilger}, {Mantovani}, {Venturi}, {Rius}, {Ros},
  {Trigilio}, \& {Whitney}}]{guirado_95}
{Guirado}, J.~C., {Marcaide}, J.~M., {Alberdi}, A., {et~al.} 1995, \aj, 110,
  2586

\bibitem[{{Hada} {et~al.}(2011){Hada}, {Doi}, {Kino}, {Nagai}, {Hagiwara}, \&
  {Kawaguchi}}]{hada_11}
{Hada}, K., {Doi}, A., {Kino}, M., {et~al.} 2011, Nat, 477, 185

\bibitem[{{Honma} {et~al.}(2008){Honma}, {Tamura}, \& {Reid}}]{honma_08}
{Honma}, M., {Tamura}, Y., \& {Reid}, M.~J. 2008, \pasj, 60, 951

\bibitem[{{Hovatta} {et~al.}(2014){Hovatta}, {Aller}, {Aller}, {Clausen-Brown},
  {Homan}, {Kovalev}, {Lister}, {Pushkarev}, \& {Savolainen}}]{hovatta_14}
{Hovatta}, T., {Aller}, M.~F., {Aller}, H.~D., {et~al.} 2014, \aj, 147, 143

\bibitem[{{Jorstad} \& {Marscher}(2016)}]{JorstadMarscher2016}
{Jorstad}, S., \& {Marscher}, A. 2016, Galaxies, 4, 47

\bibitem[{{Jorstad} {et~al.}(2013){Jorstad}, {Marscher}, {Smith}, {Larionov},
  {Agudo}, {Gurwell}, {Wehrle}, {L{\"a}hteenm{\"a}ki}, {Nikolashvili},
  {Schmidt}, {Arkharov}, {Blinov}, {Blumenthal}, {Casadio}, {Chigladze},
  {Efimova}, {Eggen}, {G{\'o}mez}, {Grupe}, {Hagen-Thorn}, {Joshi},
  {Kimeridze}, {Konstantinova}, {Kopatskaya}, {Kurtanidze}, {Kurtanidze},
  {Larionova}, {Larionova}, {Sigua}, {MacDonald}, {Maune}, {McHardy}, {Miller},
  {Molina}, {Morozova}, {Scott}, {Taylor}, {Tornikoski}, {Troitsky}, {Thum},
  {Walker}, {Williamson}, {Sallum}, {Consiglio}, \&
  {Strelnitski}}]{2013ApJ...773..147J}
{Jorstad}, S.~G., {Marscher}, A.~P., {Smith}, P.~S., {et~al.} 2013, \apj, 773,
  147

\bibitem[{{K\"onigl}(1981)}]{konigl_81}
{K\"onigl}, A. 1981, \apj, 243, 700

\bibitem[{{Kovalev} {et~al.}(2008){Kovalev}, {Lobanov}, {Pushkarev}, \&
  {Zensus}}]{kovalev_08}
{Kovalev}, Y.~Y., {Lobanov}, A.~P., {Pushkarev}, A.~B., \& {Zensus}, J.~A.
  2008, \aap, 483, 759

\bibitem[{{Kutkin} {et~al.}(2014){Kutkin}, {Sokolovsky}, {Lisakov}, {Kovalev},
  {Savolainen}, {Voytsik}, {Lobanov}, {Aller}, {Aller}, {Lahteenmaki},
  {Tornikoski}, {Volvach}, \& {Volvach}}]{kutkin_14}
{Kutkin}, A.~M., {Sokolovsky}, K.~V., {Lisakov}, M.~M., {et~al.} 2014, \mnras,
  437, 3396

\bibitem[{{Lara} {et~al.}(1994){Lara}, {Alberdi}, {Marcaide}, \&
  {Muxlow}}]{lara_94}
{Lara}, L., {Alberdi}, A., {Marcaide}, J.~M., \& {Muxlow}, T.~W.~B. 1994, \aap,
  285, 393

\bibitem[{{Lobanov}(1998)}]{lob_98}
{Lobanov}, A.~P. 1998, \aap, 330, 79

\bibitem[{Marscher(2009)}]{Marscher:2009vs}
Marscher, A.~P. 2009, Approaching Micro-Arcsecond Resolution with VSOP-2:
  Astrophysics and Technologies ASP Conference Series, 402, 194

\bibitem[{Marscher(2012)}]{Marscher:2012gh}
---. 2012, Int. J. Mod. Phys. Conf. Ser., 08, 151

\bibitem[{{Marscher} {et~al.}(2002){Marscher}, {Jorstad}, {G{\'o}mez}, {Aller},
  {Ter{\"a}sranta}, {Lister}, \& {Stirling}}]{marscher_02}
{Marscher}, A.~P., {Jorstad}, S.~G., {G{\'o}mez}, J.-L., {et~al.} 2002, \nat,
  417, 625

\bibitem[{{Marscher} {et~al.}(2008){Marscher}, {Jorstad}, {D'Arcangelo},
  {Smith}, {Williams}, {Larionov}, {Oh}, {Olmstead}, {Aller}, {Aller},
  {McHardy}, {L{\"a}hteenm{\"a}ki}, {Tornikoski}, {Valtaoja}, {Hagen-Thorn},
  {Kopatskaya}, {Gear}, {Tosti}, {Kurtanidze}, {Nikolashvili}, {Sigua},
  {Miller}, \& {Ryle}}]{marscher_08}
{Marscher}, A.~P., {Jorstad}, S.~G., {D'Arcangelo}, F.~D., {et~al.} 2008, \nat,
  452, 966

\bibitem[{{Marscher} {et~al.}(2010){Marscher}, {Jorstad}, {Larionov}, {Aller},
  {Aller}, {L{\"a}hteenm{\"a}ki}, {Agudo}, {Smith}, {Gurwell}, {Hagen-Thorn},
  {Konstantinova}, {Larionova}, {Larionova}, {Melnichuk}, {Blinov},
  {Kopatskaya}, {Troitsky}, {Tornikoski}, {Hovatta}, {Schmidt}, {D'Arcangelo},
  {Bhattarai}, {Taylor}, {Olmstead}, {Manne-Nicholas}, {Roca-Sogorb},
  {G{\'o}mez}, {McHardy}, {Kurtanidze}, {Nikolashvili}, {Kimeridze}, \&
  {Sigua}}]{marscher_10}
{Marscher}, A.~P., {Jorstad}, S.~G., {Larionov}, V.~M., {et~al.} 2010, \apjl,
  710, L126

\bibitem[{{Mart{\'{\i}}} {et~al.}(2016){Mart{\'{\i}}}, {Perucho}, \&
  {G{\'o}mez}}]{Marti2016}
{Mart{\'{\i}}}, J.~M., {Perucho}, M., \& {G{\'o}mez}, J.~L. 2016, ArXiv
  e-prints

\bibitem[{{Mart{\'{\i}}-Vidal} {et~al.}(2011){Mart{\'{\i}}-Vidal}, {Marcaide},
  {Alberdi}, {P{\'e}rez-Torres}, {Ros}, \& {Guirado}}]{martividal_11}
{Mart{\'{\i}}-Vidal}, I., {Marcaide}, J.~M., {Alberdi}, A., {et~al.} 2011,
  \aap, 533, A111

\bibitem[{{Mizuno} {et~al.}(2015){Mizuno}, {G{\'o}mez}, {Nishikawa}, {Meli},
  {Hardee}, \& {Rezzolla}}]{2015ApJ...809...38M}
{Mizuno}, Y., {G{\'o}mez}, J.~L., {Nishikawa}, K.-I., {et~al.} 2015, \apj, 809,
  38

\bibitem[{Molina(2016)}]{molina_bllac}
Molina, S. 2016, \aj

\bibitem[{{O'Sullivan} \& {Gabuzda}(2009)}]{osullivan_09}
{O'Sullivan}, S.~P., \& {Gabuzda}, D.~C. 2009, \mnras, 400, 26

\bibitem[{{Perucho} {et~al.}(2010){Perucho}, {Bosch-Ramon}, \&
  {Khangulyan}}]{perucho_10}
{Perucho}, M., {Bosch-Ramon}, V., \& {Khangulyan}, D. 2010, \aap, 512, L4

\bibitem[{{Reid} \& {Honma}(2014)}]{reid_micro}
{Reid}, M.~J., \& {Honma}, M. 2014, \araa, 52, 339

\bibitem[{{Rioja} \& {Dodson}(2011)}]{rioja_11a}
{Rioja}, M., \& {Dodson}, R. 2011, AJ, 141, 114

\bibitem[{{Rioja} {et~al.}(2011){Rioja}, {Dodson}, {Malarecki}, \&
  {Asaki}}]{rioja_11b}
{Rioja}, M., {Dodson}, R., {Malarecki}, J., \& {Asaki}, Y. 2011, AJ, 142, 157

\bibitem[{{Rioja} {et~al.}(2016){Rioja}, {Dodson}, {Orosz}, {Imai}, \&
  {Frey}}]{rioja_16}
{Rioja}, M., {Dodson}, R., {Orosz}, G., {Imai}, H., \& {Frey}, S. 2016, AJ

\bibitem[{{Rioja} {et~al.}(2009){Rioja}, {Dodson}, {Porcas}, {Ferris},
  {Reynolds}, {Sasao}, \& {Schilizzi}}]{rioja_09}
{Rioja}, M., {Dodson}, R., {Porcas}, R.~W., {et~al.} 2009, in 8th International
  e-VLBI Workshop, 14

\bibitem[{{Rioja} {et~al.}(2015){Rioja}, {Dodson}, {Jung}, \&
  {Sohn}}]{rioja_15}
{Rioja}, M.~J., {Dodson}, R., {Jung}, T., \& {Sohn}, B.~W. 2015, \aj, 150, 202

\bibitem[{{Rioja} \& {Porcas}(1998)}]{rioja_98}
{Rioja}, M.~J., \& {Porcas}, R.~W. 1998, in Astronomical Society of the Pacific
  Conference Series, Vol. 144, IAU Colloq. 164: Radio Emission from Galactic
  and Extragalactic Compact Sources, ed. J.~A. {Zensus}, G.~B. {Taylor}, \&
  J.~M. {Wrobel}, 95

\bibitem[{{Rioja} {et~al.}(2014){Rioja}, {Dodson}, {Jung}, {Sohn}, {Byun},
  {Agudo}, {Cho}, {Lee}, {Kim}, {Kim}, {Oh}, {Han}, {Je}, {Chung}, {Wi},
  {Kang}, {Lee}, {Chung}, {Ryoung Kim}, {Kim}, {Lee}, {Roh}, {Oh}, {Yeom},
  {Song}, \& {Kang}}]{rioja_14}
{Rioja}, M.~J., {Dodson}, R., {Jung}, T., {et~al.} 2014, \aj, 148, 84

\bibitem[{{Ros} {et~al.}(2001){Ros}, {Marcaide}, {Guirado}, \&
  {P{\'e}rez-Torres}}]{ros_01}
{Ros}, E., {Marcaide}, J.~M., {Guirado}, J.~C., \& {P{\'e}rez-Torres}, M.~A.
  2001, \aap, 376, 1090

\bibitem[{{Sokolovsky} {et~al.}(2011){Sokolovsky}, {Kovalev}, {Pushkarev},
  {Mimica}, \& {Perucho}}]{sokolovsky_11}
{Sokolovsky}, K.~V., {Kovalev}, Y.~Y., {Pushkarev}, A.~B., {Mimica}, P., \&
  {Perucho}, M. 2011, \aap, 535, A24

\bibitem[{Sovers {et~al.}(1998)Sovers, Fanselow, \& Jacobs}]{new_geodetic}
Sovers, O.~J., Fanselow, J.~L., \& Jacobs, C.~S. 1998, Rev. Mod. Phys., 70,
  1393

\bibitem[{Williams {et~al.}(1979)Williams, Nixon, Reilly, Withington, \&
  Bathker}]{sxfeeds}
Williams, W., Nixon, D., Reilly, H., Withington, J., \& Bathker, D. 1979, DSN
  Progress Report 42, 52, 51

\end{thebibliography}

\end{document}